\documentclass[12pt,preprint]{aastex6}

\usepackage{amsmath}
\usepackage{natbib}
\usepackage{xcolor}
\usepackage[]{graphics}
\usepackage{epsfig}
\usepackage{epstopdf}

\usepackage{mathrsfs}
\usepackage{appendix}

\begin{document}

\title{Observational Constraints on the Rastall gravity from Rotation Curves of Low Surface Brightness Galaxies}

\author{
   Meirong Tang \altaffilmark{1,3,4,5}
   Zhaoyi Xu \altaffilmark{2}
  Jiancheng Wang \altaffilmark{1,3,4,5}
 }

\altaffiltext{1}{Yunnan Observatories, Chinese Academy of Sciences, 396 Yangfangwang, Guandu District, Kunming, 650216, P. R. China; {\tt  mrtang@ynao.ac.cn, xuzy@ihep.ac.cn,  jcwang@ynao.ac.cn
}}
\altaffiltext{2}{Key Laboratory of Particle Astrophysics, Institute of High Energy Physics, Chinese Academy of Sciences, Beijing 100049, P.R. China}
\altaffiltext{3}{Key Laboratory for the Structure and Evolution of Celestial Objects, Chinese Academy of Sciences, 396 Yangfangwang, Guandu District, Kunming, 650216, P. R. China}
\altaffiltext{4}{Center for Astronomical Mega-Science, Chinese Academy of Sciences, 20A Datun Road, Chaoyang District, Beijing, 100012, P. R. China}
\altaffiltext{5}{University of Chinese Academy of Sciences, Beijing, 100049, P. R. China}

\shorttitle{Observational Constraints for Rastall gravity}
\shortauthors{Tang et al.}

\begin{abstract}
The Rastall gravity is a modification of Einstein's general relativity, in which the energy-momentum conservation is not satisfied and depends on the gradient of the Ricci curvature. It is in dispute whether the Rastall gravity is equivalent to the general relativity (GR). In this work, we constrain the theory using the rotation curves of Low Surface Brightness (LSB) spiral galaxies. Through fitting the rotation curves of LSB galaxies, we obtain the parameter $\beta$ of the Rastall gravity. The $\beta$ values of LSB galaxies satisfy Weak Energy Condition (WEC) and Strong Energy Condition(SEC). Combining the $\beta$ values of type Ia supernovae and gravitational lensing of elliptical galaxies on the Rastall gravity, we conclude that the Rastall gravity is equivalent to the general relativity.

\end{abstract}


\section{INTRODUCTION}

One hundred years ago, Einstein proposed the general theory of relativity describing the gravity successfully. One of the important fundamental of the GR is to assume that the covariant derivative of the energy-momentum tensor is zero, and the GR naturally satisfies the equivalence principle. Rastall generalized the covariant conservation of energy-momentum tensor \citep{1972PhRvD...6.3357R, 1976CaJPh..54...66R}, and obtained the conservation equation of energy-momentum tensor with the form $T^{\mu\nu}_{;\mu}=\lambda R^{,\nu}$, where $T^{\mu\nu}$ is energy momentum tensor, $R$ is Ricci curvature (or Ricci scalar) and $\lambda$ is the parameter of the Rastall gravity. This theory can reduce to the GR in asymptotically flat space-time.
However, the Rastall gravity is still a controversial gravitational theory. One view is that the Rastall gravity is equivalent to the GR \citep{2018PhLB..782...83V}, the parameter $\lambda$ represents the re-arrangement of perfect fluid matter. According to this view, we just need to redefine the energy-momentum tensor for satisfying the covariant conservation. The only change is the addition of matter fields with different distributions in space-time. On the contrary, \cite{2018EPJC...78...25D} considered that the Rastall gravity is not equivalent to the GR.  The Rastall gravity strengthens the role of Mach principle in gravity theory \citep{2006gr.qc....10070M}, in which the local structure depends on the distribution of matter in entire space-time.

Although the nature of the Rastall gravity is not clear enough, we try to constrain its property by observational events. On the cosmological scale, \cite{2013EPJC...73.2425B} used the data of type Ia supernovae to analyse the gravity model and achieved some good results. Recently, \emph{Li and Xu et al.} \footnote{this article is being reviewed, but not on arXiv} acquired the measured value of the Rastall gravity parameter on the scale of elliptical galaxies, and supposed that the Rastall gravity can explain the mass distribution of elliptical galaxies. The measured value is consistent with one constrained by energy condition, but the result of the Rastall gravity in accordance with the GR on a large scale is the requirement of the theory itself. To test the result, we need the data of galaxies on a smaller scale, such as galaxy nucleus and spiral galaxies. In this study, we will use the data of rotation curves from low surface brightness (LSB) galaxies to constrain the Rastall gravity.

This paper is organized as follows. In Section \ref{metric}, we introduce the perfect fluid dark matter (PFDM) and Rastall gravity. In Section \ref{motion}, we constrain the Rastall gravity model from the rotation curves of LSB spiral galaxies. In Section \ref{discussion}, we discuss our results and compare them with previous results. Finally, in Section \ref{conclusion}, we present our conclusions.

\section{PERFECT FLUID DARK MATTER AND RASTALL GRAVITY}
\label{metric}

\subsection{Perfect fluid dark matter in GR}

\cite{2003CQGra..20.1187K} obtained the black hole solution of perfect fluid matter in GR. They assumed that the equation of state defined by the ratio of pressure to density of perfect fluid matter $\omega=p/\rho$ is a constant, where $p$ and $\rho$ are the pressure and density of perfect fluid matter, respectively. The expression of black hole solution is

\begin{equation}
ds^{2}=-f(r)dt^{2}+\dfrac{1}{g(r)}dr^{2}+r^{2}d\Omega^{2},
\label{BH_solution_GR}
\end{equation}
where $f(r)$ and $g(r)$ are written as

\begin{equation}
f(r)=g(r)=1-\dfrac{2M}{r}-\dfrac{\alpha}{r^{1+3\omega}}.
\label{BH1_solution_GR}
\end{equation}
$M$ represents the mass of black hole, and $\alpha$ is the intensity parameter of perfect fluid matter around black hole. If the equation of state is given by $\omega=-1/3$, the black hole solution represents a Schwarzschild black hole under the perfect fluid dark matter background \citep{2003gr.qc.....3031K, 2017PhRvD..95f4015X, 2010PhLB..694...10R, 2000astro.ph..3105G}. In general, the perfect fluid dark matter (PFDM) is quintessence matter, because only in this kind of matter, the equation of state is possibly equal to $-1/3 $. At the same time, the solution can also be understood as the black hole solution under the PFDM model. In this situation, the flatness of rotation curves of spiral galaxies at a long distance can be explained naturally. Here, we assume that this property continues to be valid in the Rastall gravity.

\subsection{Perfect fluid dark matter in Rastall gravity}

\cite{2017PhLB..771..365H} generalized \cite{2003CQGra..20.1187K} solution from the GR to the Rastall gravity, and obtained the spherically symmetric black hole solution in perfect fluid matter. This solution has a form of

\begin{equation}
f(r)=g(r)=1-\dfrac{2M}{r}-\alpha r^{-\dfrac{1+3\omega-6\beta(1+\omega)}{1-3\beta(1+\omega)}},
\label{BH1_solution_Rg}
\end{equation}
where $\kappa\lambda$ is a parameter of the Rastall gravity, which determines the distribution of perfect fluid matter. For convenience, we write $\kappa\lambda$ as $\beta$ throughout this article, i.e. $\beta=\kappa\lambda$.  For the PFDM ($\omega=-1/3$), the energy density $\rho_{DM}$ can be derived from the Einstein equation. Because the motion velocity of dark matter particle is much smaller than the speed of light, the energy density of the PFDM can approximate to the mass density. Here, from \cite{2017PhRvL.119k1102K}, the baryon matter can be treated as an index disk, i.e. $\rho_{b}=\Sigma_{0}exp[-r/r_{d}]\delta(z)$, where $\Sigma_{0}$ and $r_{d}$ are the central surface density and scale radius of the disk, respectively. In this space-time metric, using the mass density of the PFDM halo and baryon disk, we can calculate the total mass function described as $M(r) =4\pi\int_{0}^{r}\rho_{DM}r^{2}dr+2\pi\int_{0}^{r}\rho_{b}rdr$. We then obtain the rotation velocity of stars on the equatorial plane written as \citep{2018EPJC...78..513X}

\begin{equation}
\upsilon(r) =\sqrt{\dfrac{GM(r)}{r}}
              =\sqrt{\dfrac{G\alpha}{2}\dfrac{1-4\beta}{1-2\beta}r^{\dfrac{4\beta}{1-2\beta}}-2\pi G\Sigma_{0}r_{d}exp[\dfrac{-r}{r_{d}}](1+\dfrac{r_{d}}{r})} .
\label{rotation_velocity}
\end{equation}
where G is the gravitational constant. We will use this equation to fit the rotation curves of LSB galaxies, and get the $\beta$ values of the Rastall gravity. The parameter $r_{d}$ is $2 kpc$ \citep{2017PhRvL.119k1102K} in section 3.

\subsection{Energy condition in Rastall gravity}

In the theory of gravity, it is extremely difficult to solve the equation of gravitational field. Through the Einstein field equation, we can know that the distribution of energy-momentum tensor determines the structure of space-time. Due to the complexity of matter distribution, the energy- momentum tensor can not be expressed by a specific form. Therefore, the certain conditions that the energy density is greater than or equal to zero were used to study the gravitational field equation.

In 1955, Raychaudhuri formally proposed the basic equation of energy condition, such as weak energy condition and strong energy condition. Under these energy conditions, the fundamental properties of gravity are satisfied. In the references \cite{2017PhLB..771..365H} and \cite{2018EPJC...78..513X}, they made a specific study on the energy conditions of the Rastall gravity, and found that the constraint of weak energy condition and strong energy condition on the Rastall parameter $\beta$ is equal in the assumption of perfect fluid. They can be given by

\begin{equation}
(3\beta(1+\omega)-3\omega)(1-4\beta)\geq0.
\label{WEC_SEC}
\end{equation}

If $\omega=-1/3$, the perfect fluid matter is described by the PFDM model, and the range of the Rastall parameter $\beta$ is $-1/2<\beta<1/4$. If the $\beta$ obtained by fitting the observation data is within this range and is a constant on the scale of spiral galaxy, elliptical galaxy and cosmology, the Rastall gravity will be supported. In contrast, the model will be excluded.

\section{CONSTRAINTS FROM ROTATION CURVES OF LSB GALAXIES}
\label{motion}

In this section,  according to the Eq.(\ref{rotation_velocity}), we adopt the Bayesian method \citep{2010arXiv1008.4686H} to fit the rotation curves of 16 LSB spiral galaxies, and obtain the good fits overall, with $\chi^{2}/dof<1$ for 15 galaxies ( F563-1,  F568-3, F583-1, F571-8, F579-v1, F583-4,  F730-v1, U5750, U11454, U11616, U11648, U11819, ESO0140040,  ESO2060140, ESO3020120 ), and $\chi^{2}/dof<2$ for one galaxy ( ESO4250180 ). Here, the predicted velocity $\upsilon_{pre}$ is from Eq.(\ref{rotation_velocity}) as $\upsilon(r)$, and the observed velocity $\upsilon_{obs}$ is from an astronomical website. For each galaxy, we assume that it has $i$ data points. Therefore, the likelihood function can be expressed as

\begin{equation}
ln\mathcal{L}=-\dfrac{1}{2}\sum\limits_{i= 0}^{16}[\dfrac{(\upsilon_{pre}^{i}-\upsilon_{obs}^{i})^{2}}{s_{i}^{2}}+ln(2\pi s_{i}^{2})],
\label{likelihood_function}
\end{equation}
where
\begin{equation}
s_{i}^{2}=\delta^{2}+(\upsilon_{err}^{i})^{2},
\label{likelihood1_function}
\end{equation}
and $\delta$ is the intrinsic scatter between $\upsilon_{pre}$ and  $\upsilon_{obs}$, which is considered as a free parameter in our Bayesian analysis. $\upsilon_{err}$ is the measurement error of $\upsilon_{obs}$. Now, the posterior probability function is written as

\begin{equation}
p(\alpha, \beta, \Sigma_{0}, \delta|{\upsilon_{obs}})=\mathcal{L}({\upsilon_{obs}|\alpha, \beta, \Sigma_{0}, \delta})p(\alpha, \beta, \Sigma_{0}, \delta).
\label{posterior_probability}
\end{equation}

Here, for each LSB galaxy, we choose a flat prior $p(\alpha, \beta, \Sigma_{0}, \delta)$, and use the Python implementation named Emcee \citep{2013PASP..125..306F} along with four free parameters $\alpha, \beta, \Sigma_{0}, \delta$ to fit the rotation curves. Our results are shown in Table 1. 

\begin{table*}
\centering
  \noindent Table 1. Best fitting results from the rotation curves of 16 LSB spiral galaxies using the eq.(\ref{rotation_velocity}). 
Column (1), (2) and (3 ) are the name of galaxy, the fitting values of Rastall parameter $\beta$ and the $\chi^{2}$ values, respectively.\\
  [2mm]
   \begin{scriptsize}{
  \begin{tabular}{lll}
  \hline \hline
 Galaxy  & $\beta$  & $\chi^{2}/dof$ \\
 (1)  & (2) & (3) \\
  \hline
    F563-1 &  0.054    &  0.879   \\
    F568-3 &  0.154   &  0.866  \\
    F583-1 &  0.15      &  0.836   \\
    F571-8 &  0.143       &  0.882   \\
    F579-v1 &  0.048      &  0.13   \\
    F583-4 &  0.141        &  0.211   \\
    F730-v1 &  0.095    &  0.53   \\
    U5750  &  0.146     &  0.82   \\
    U11454 &  0.118    &  0.826   \\
    U11616 &  0.122        &  0.808   \\
    U11648 &  0.132      &  0.356   \\
    U11819 &  0.147      &  0.958   \\
    ESO0140040 & 0.083     &  0.757   \\
    ESO2060140 &  0.1        &  0.812   \\
    ESO3020120 &  0.135     &  0.397  \\
    ESO4250180 &  0.121     &  1.675   \\
   \hline \\
  \end{tabular}
  }
  \end{scriptsize}
\end{table*}

\section{DISCUSSION}
\label{discussion}

In this study, we constrain the parameter $\beta$ of the Rastall gravity by fitting the data of the rotation curves from 16 LSB spiral galaxies. Comparing the previous results of type Ia supernovae and elliptical galaxies, we support that the Rastall gravity is equivalent to the GR.

On the scale of spiral galaxy, the values of parameter $\beta$ we obtained are of the order of $10^{-1}$ and within the limits of strong energy condition in the Rastall gravity.
On the cosmological scale, \cite{2013EPJC...73.2425B} used the data of type Ia supernovae to constrain the parameter $\beta$, and found that the $\beta$ value is of the order of $10^{-4}$. This is inconsistent with our results by three orders of magnitude. If our analysis is correct, then the value of the Rastall parameter will cause some troubles. For example, the value of the parameter $\beta$ on the scale of spiral galaxy is not same as the value on the cosmological scale, which indicates that the parameter $\beta$ is not universally applicable. On the other hand, \emph{Li and Xu et al.} recently used the gravitational Lensing data of elliptical galaxies to constrain the Rastall parameter. Their $\beta$ values obtained by fitting the observed data are consistent with ours, which implies that the parameter values of the Rastall gravity are consistent between elliptical galaxies and spiral galaxies. At the same time, according to the Rastall gravity model, the result of the Rastall gravity in accordance with the GR on a large scale is the requirement of the theory itself, and the difference only appears on a smaller scale. Our results will be more convincing.

The nature of the Rastall gravity is in dispute. What is the explanation of our results for this? If the Rastall gravity is equivalent to the GR, the $\beta$ only represents the distribution of matter, and the different systems will have different values, therefore it is easy to understand the difference of the $\beta$ between the galactic scale and the cosmological scale. If the Rastall gravity is not equivalent to the GR, on the basis of the assumption of the Rastall gravity, the $\beta$ values should be same for the galactic scale and the cosmological scale, but this is contrary to our analysis results. So we will conclude that the Rastall parameter $\beta$ can only be understood as the parameter determining the distribution of matter in space-time. Thus our results support that the Rastall gravity is equivalent to the GR.

\section{CONCLUSIONS}
\label{conclusion}

In this work, using the rotation curves of 16 LSB spiral galaxies, we obtained the values of parameter $\beta$ in the Rastall gravity model. These values are about 0.1 and satisfy the strong energy condition. And after comparing the results of type Ia supernovae and elliptical galaxies, we found that our results support that the Rastall gravity is equivalent to the GR, and then the values of parameter $\beta$ can only be understood as the re-arrangement of matter in space-time, the difference between the galactic scale and the cosmological scale can be easily explained.

\acknowledgments
We thank Kuzio de Naray, R., McGaugh, S.S., and de Blok for providing us the rotation curve data and thank Li Rui for his help in programming. We acknowledge the anonymous referee for a constructive report that has significantly improved this paper. We acknowledge the financial support from the National Natural Science Foundation of China under grants No. 11503078, 11573060 and 11661161010.


\end{document}